\documentclass[a4paper]{article}

\usepackage{icrc2013}

\title{Observations of VHE gamma-ray binaries with the MAGIC Telescopes}

\shorttitle{Observations of VHE gamma-ray binaries with MAGIC}

\authors{
A. L\'opez-Oramas$^{1}$,
O.Blanch Bigas$^{1}$,
J.Cortina$^{1}$,
D.Hadasch$^{2}$,
A.Herrero$^{3,4}$,
B. Marcote$^{5}$,
P.Munar-Adrover$^{5}$,
J.Mold\'on$^{5,6}$,
J.M.Paredes$^{5}$,
I.Ribas$^{5}$,
M.Rib\'o$^{5}$,
D.Torres$^{2}$,
R.Zanin$^{5}$
for the MAGIC Collaboration AND.
J. Casares$^{3,4}$,
N.Rea$^{2}$

}

\afiliations{
$^1$ Institut de F\'isica d' Altes Energ\`ies (IFAE), Universitat Aut\`onoma de Barcelona, Edifici Cn, 08193 Bellaterra, Barcelona  \\
$^2$IEEC-CSIC, Campus UAB Torre C5 Parell 2a planta Barcelona 08193\\
$^3$ Instituto de Astrof\'isica de Canarias (IAC), 38200, La Laguan, Tenerife, Spain\\
$^4$ Departamento de Astrofísica, Universidad de La Laguna, Avda. Asdtrof\'isico Francisco S\'anchez s/n E-38271, La Laguna, Tenerife, Spain\\
$^5$ Universitat de Barcelona, Institut de Ci\`encies del Cosmos (IEEC-UB) \\
$^6$ ASTRON Netherlands Institute for Radio Astronomy, Oude Hoogeveensedijk 4, 7991 PD Dwingeloo, The Netherlands

}

\email{alopez@ifae.es}

\abstract{Several binary systems, composed of a star and a compact object, have been detected in the GeV-TeV range. Several systems have been
observed but only a handful of sources have shown emission at those
energies. Here, we present the observations conducted by MAGIC
of different $\gamma$-ray binary systems. On one hand, we show the
latest studies on the binary system LS I +61 303, which displays
variability on different timescales. With the latest MAGIC observations,
we will try to shed light on our understanding of this source, by
presenting super-orbital and multi-wavelength studies. On the other hand,
we show the observational results on the binary system HD 215227. This 
source has been proposed as a new $\gamma$-ray binary for being
spatially coincident with the gamma-ray source AGL J2241+4454 detected by
AGILE at E \textgreater 100 GeV.}


\begin{document}
\maketitle

\section{Introduction}
Some binary systems are composed of a star and a compact object, which can be either a neutron star (NS) or a stellar-mass black hole (BH). If the companion star is a young star of spectral type O or B, there will be a mass transfer to the compact object via decretion disk (for fast rotators), strong stellar wind (for luminous objects) or Roche lobe overflow. The accreted material will heat up and produce X-ray emission. These systems are known as High Mass X-ray Binaries (HMXRBs). Only a handful of these bodies have been detected up to GeV-TeV energies, the \emph{gamma-ray binaries}, and most of their non-thermal luminosity is emitted in the high energy and very high energy (VHE) gamma-ray band.\\

The first TeV binary system detected was PSR B1259-63, a radio pulsar orbiting a Be star \cite{bib:Aharonian2005}. Later, other binaries have been detected at VHE energies, as LS I +61 303, LS 5039 or HESS J0632+057. In this proceeding we will show the latest results obtained by MAGIC in the field of VHE gamma-ray binaries.


\section{LS I + 61 303}

\subsection{Unveiling the system}
LS I +61 303 is a binary system composed of a Be star (spectral type B0 Ve) with a circumstellar disk and a compact object of unknown nature. It is located at a distance of $\sim$ 2kpc. The binary is eccentric (e = 0.55 $\pm$ 0.05) and has an orbital period of 26.4960 $\pm$ 0.0028 days, which was determined by radio observations \cite{bib:Gregory}. The periastron passage occurs at phase $\phi$=0.23-0.3, depending the orbital solution (\cite{bib:Gregory}, \cite{bib:Grundstrom},\cite{bib:Aragona}). Other orbital parameters remain poorly known and present large uncertainties. LS I +61 303 was first detected in the VHE gamma-ray band by MAGIC in 2006 \cite{bib:magic2006}\\

The system has been observed at different wavelengths: IR, radio, X-rays and HE and VHE gamma rays. It presents periodic radio outburst which are widely associated to the orbital period. Radio outburst start from phase 0.45 and last until phase 0.95 \cite{bib:paredes}. This periodical modulation is also present in other wavelengths. In X-rays the outburst is visible between phases $\phi$ = 0.4 -0.8 \cite{bib:xrays}. The binary also shows a super orbital modulation with periodicity of about 1667 $\pm$ 8 days \cite{bib:Gregory}, which leads to soft and hard states in both wavelengths. The source has been observed also at HE gamma rays by the Fermi telescope, and the outburst was exhibited at phases 0.3-0.45, just after the periastron passage.\\

The VHE emission component of LS I +61 303 is modulated with the 26.6 $\pm$ 0.2 days orbital period \cite{bib:albert2009}. A TeV peak was first detected at phases  $\phi$ =0.6-0.7 at a $\sim$ 16 $\%$ Crab Nebula flux and presenting a Crab-like spectrum with spectral index $\alpha$$\sim$ 2.6. The fact that the HE and VHE outburst does not occur at the same phases may indicate that the processes which generate these two components might be different. Despite a periodic outburst is almost always present in this phase range, emission can also be detected among phases  $\phi$ =0.8-1. The emission in this range does not present orbital modulation. No emission around periastron has been detected by the MAGIC telescopes, although VERITAS detected the system at near-periastron phases during fall 2010 \cite{bib:acciari2011}.\\

In the winter 2009-2010 campaign performed by MAGIC, the source was detected in a low-state flux emission level \cite{bib:lowemission}. The integral flux above 300 GeV was (6.1 $\pm$ 1.4$_{stat}$ $\pm$ 1.8$_{sys}$) x 10$^{-12}$ cm$^{-2}$ s$^{-1}$, corresponding to a 5.4 $\%$ of Crab Nebula flux, about a factor three lower than previous campaigns. Nevertheless, the TeV peak is still detected in the same phases $\phi$ = 0.6-0.7 . In these observations, the spectral fit  parameters agree with previous values reported by MAGIC.\\

Different scenarios have been proposed to explain the non-thermal emission  observed in LS I +61 303. The first scenario is the microquasar scenario \cite{bib:Romero}, where the compact object is a black hole. In this case, the companion star losses mass into an accretion disk around the black hole. Detection of extended jet-like radio-emitting structures has been interpreted as a possible evidence of the microquasar nature of the system. The second proposed scenario is the rotational powered pulsar scheme \cite{bib:Dubus}, where the companion is a neutron star. Nonetheless, none of these scenarios could be validated, due to the lack of accretion disk features like black body component in the X-ray spectrum and due to the absence of pulsations at any wavelength. A third theory  propose LS I +61 303 to be the first magnetar detected in a binary system \cite{bib:torres}. A flip-flop magnetar model has been proposed to explain the orbital and super orbital emissions of the binary, where it would change from a rotational powered regime (in apastron), where the shocked wind would produce the TeV emission, to a propeller regime (in periastron) where particles are accelerated only until sub-TeV energies. This model explains the anticorrelation at HE and VHE energies and also the low flux emission state.\\

\subsection{Observations}

Observations were performed with the MAGIC telescopes, which is an stereoscopic system composed of two 17 m diameter air imaging Cherenkov telescopes placed in the Canary island of La Palma, at 2225 m a.s.l. The sensitivity of the system is 5$\sigma$ signal above 300 GeV for a source emitting (0.76$\pm0.03$)$\%$ of Crab Nebula flux in 50 hours. The analysis has been performed with MAGIC standard reconstruction software.\\

Two observation campaigns were held. On the first one, the binary system was observed between August 2010 and January 2011 under bright moon conditions, covering seven orbital periods and phases from 0.4 to 1. After a tight pre-selection, about $\sim$ 30 hours of good data remained. On the second campaign, a multiwavelength observation was performed, in order to search for correlation/anticorrelation between TeV emission and the mass loss rate of the Be star, which will allow to test the model presented in \cite{bib:torres} In this case, the observations covered the phase range between 0.7 and 1, due to observability issues. Furthermore, this phase range exhibits more variability in the flux of the binary system: the source emits periodic outburst in the 0.6-0.7, but sometimes it also shows emission in the 0.8-0.9 phase range, but not periodically. This range presents the highest variability period to period, which makes it suitable to study correlations with other wavelengths. Nightly observations of about 2.5 h each with MAGIC were performed, to ensure to detect the system even at relatively low emission level with good statistics. 

\subsection{Results}

The overall significance for the cycle from August 2010 and January 2011 is above 10 sigma. Most of the emission comes from January 2011 data, where the source was detected at phases 0.5-0.7. These results reveal that LS I +61 303 recovered its high energy state. These observations were performed at a super orbital phase of $\phi$$_{suporb}$ $\sim$0.32, which follows the predictions made by \cite{bib:jianli}.\\

\section{HD 215227}

\subsection{A new gamma-ray binary?}
On July 27th, 2010, AGILE detected a gamma-ray signal above 100 MeV from an unidentified point-like source, AGL J2241+4454 \cite{bib:atel}. The detection showed a significance $>$5$\sigma$ and a flux of 10$^{-8}$ ph cm$^{-2}$ s$^{-1}$.\\

Lying within the error circle of the satellite, there is a Be star, HD 215227, which was proposed to be the optical counterpart of the new gamma-ray source \cite{bib:williams}. Optical spectroscopy of this star confirmed its binary nature \cite{bib:casaresHD}. It seems that the Be star exhibits a circumstellar disk and that the optical emission is modulated. This modulation is an indicator of instabilities in the circumstellar disk. The binary has an orbital period of 60.37 $\pm$  0.04 days and it is located at 2.6 $\pm$ 1.0 kpc \cite{bib:williams}. 

\subsection{Observations}
The observations of HD 215227 were performed between 23rd May and 19th June 2012 using the MAGIC telescopes  with the aim of detecting the source for the first time at VHE gamma rays. Observations were performed in mono mode, as one of the telescopes was not operational. The sensitivity for mono observations above 100 GeV is about 2.5\% of the Crab Nebula flux in 50 hours.\\

It was observed for a zenith angle range between 22$^\circ$ and 51$^\circ$. The observations covered the phases 0.08 - 0.53 (post-periastron), because they were the only phases visible from the MAGIC site on that moment. This range covers the periastron passage and also phases where other binaries exhibit they outbursts at VHE. \\

\subsection{Results}
After a pre-selection of good-quality data, $\sim$23.5 h of effective time remained. No signal was detected from the whole data sample. Total significance is about $\sim$0.5$\sigma$. A nightly analysis neither revealed a hint of signal. The integral flux upper limit set for the whole data set is:\\

\begin{equation}
F(E>300 GeV)=2.11\cdot10^{-12}cm^{-2} s^{-1}\\
\end{equation}

\begin{figure}[th!]
\centering
\includegraphics[width=0.5\textwidth]{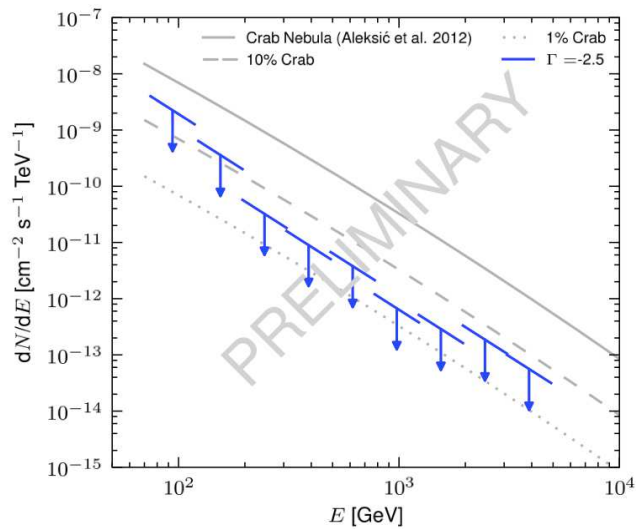}
\caption{Upper limits obtained from HD215227 MAGIC observations. The differential upper limits are calculated, assuming a spectrum described by a simple power law with spectral index -2.5, similar to other gamma-ray binaries like LS I +61 303.}
\label{icrc2013-0459-01}
\end{figure}

Differential upper limits (at 96 $\%$ confidence level) were calculated for the spectrum (see Fig.~\ref{icrc2013-0459-01}) .

\section{Conclusions}
MAGIC has observed two different gamma ray binary systems, LS I +61 303 and HD215227, which was proposed as the optical counterpart of AGL J2241+4454. LS I +61 303 observations unveiled the recovery of the high state emission of this source, fact that follows the predictions of the existence of a super-orbital modulation of about $\sim$4.5 years also in VHE gamma rays. HD 215227 was observed after the periastron passage (next to phases where AGILE reported emission) and at later phases. It showed no excess of VHE gamma rays. Upper limits have been set for the VHE emission.


\section{Acknowledgments}
We would like to thank the Instituto de Astrof\'{\i}sica de
Canarias for the excellent working conditions at the
Observatorio del Roque de los Muchachos in La Palma.
The support of the German BMBF and MPG, the Italian INFN, 
the Swiss National Fund SNF, and the Spanish MICINN is 
gratefully acknowledged. This work was also supported by the CPAN CSD2007-00042 and MultiDark
CSD2009-00064 projects of the Spanish Consolider-Ingenio 2010
programme, by grant 127740 of 
the Academy of Finland,
by the DFG Cluster of Excellence ``Origin and Structure of the 
Universe'', by the DFG Collaborative Research Centers SFB823/C4 and SFB876/C3,
and by the Polish MNiSzW grant 745/N-HESS-MAGIC/2010/0. JC acknowledges support by the Spanish Ministerio de Econom\'ia y Competitividad (MINECO) under grant AYA2010-18080.


\begin{thebibliography}{}

\bibitem{bib:Aharonian2005} Aharonian et al, 2005, A\&A, 242,1.

\bibitem{bib:Gregory} Gregory P.C., ApJ 575, 427. (2002)

\bibitem{bib:Grundstrom} Grundstrom, E. D. et al., ApJ, 656, 437 (2007)

\bibitem{bib:Aragona} Aragona et al,  ApJ 698, 514 (2009)

\bibitem{bib:magic2006} Albert et al, Science 312, 1771 (2006)

\bibitem{bib:paredes} Paredes et al, A\&A, 232, 377 (1990)

\bibitem{bib:xrays} Anderhub et al, ApJ, 706, L27  (2009)

\bibitem{bib:albert2009} Albert et al, ApJ, 693, 303 (2009)

\bibitem{bib:acciari2011} Acciari et al, ApJ, 738:3  (2011)

\bibitem{bib:lowemission} Aleksi$\acute{c}$ et al, ApJ, 746, 80 (2012)

\bibitem{bib:Romero} Romero et al, ApJ 632, 1093 (2005)

\bibitem{bib:Dubus} Dubus G., 2006, A\&A 456, 801

\bibitem{bib:torres} Torres et al, ApJ  744 (2012) 106

\bibitem{bib:jianli} Li et al, ApJ, 744:L13 (5pp)  (2012)

\bibitem{bib:atel} Lucarelli et al, 2010, ATel,2761

\bibitem{bib:williams} Williams et al, ApJ, 723, L93 (2010)

\bibitem{bib:casaresHD} Casares et al, Monthly Notices of the Royal Astronomical Society vol. 421, pag. 1103 (2012)

\end{thebibliography}
\end{document}